\documentclass[12pt]{iopart}

\usepackage{iopams}
\usepackage{graphicx}
\usepackage{color}
\usepackage{epsfig}

\begin{document}

\title[Two-band conductivity of a FeSe0.5Te0.5 film]{Two-band conductivity of a FeSe$_{0.5}$Te$_{0.5}$ film by reflectance measurements in the terahertz and infrared range}

\author{A Perucchi$^{1}$, B Joseph$^2$\footnote{Present address: 
Elettra Sincrotrone Trieste, Area Science Park, I-34012 Trieste, Italy}, S Caramazza$^2$, M Autore$^2$, E Bellingeri$^{3}$, 
S Kawale$^3$, C Ferdeghini$^{3}$, M Putti$^4$, S Lupi$^5$ and P Dore$^6$}  
\address{$^1$ INSTM Udr Trieste-ST and Elettra Sincrotrone Trieste, Area Science Park, I-34012 Trieste, Italy} 
\address{$^2$ Dipartimento di Fisica, Universit\`a di Roma Sapienza, P.le Aldo 
Moro 2, I-00185 Rome, Italy} 
\address{$^3$ CNR-SPIN, Corso Perrone 24, I-16152 Genova, Italy}
\address{$^4$ CNR-SPIN and Dipartimento di Fisica, Universit\`a di Genova,Via Dodecaneso 33, I-16146 Genova, Italy}
\address{$^5$ CNR-IOM and Dipartimento di Fisica, Universit\`a di Roma Sapienza, P.le Aldo Moro 2, I-00185 Roma, Italy}
\address{$^6$ CNR-SPIN and Dipartimento di Fisica, Universit\`a di Roma Sapienza, P.le Aldo Moro 2, I-00185 Roma, Italy}
\ead{paolo.dore@roma1.infn.it}

\begin{abstract}
We report an infrared spectroscopy study  of a 200 nm thick
FeSe$_{0.5}$Te$_{0.5}$ film grown on LaAlO$_3$ with T$_c$=13.7 K. 
We analyze the 20 K normal state absolute reflectance R$_N$
measured over a broad infrared range and the reflectance ratio 
R$_S$/R$_N$, R$_S$ being the superconducting state reflectance, 
measured at 6 K in the terahertz range down to 12 cm$^{-1}$. 
We show that the normal state model conductivity is given by two 
Drude components, one of which much broader and intense than the other.
In the superconducting state, we find that a gap 
$\Delta$=37$\pm$3 cm$^{-1}$ 
opens up in the narrow Drude band only, 
while the broad Drude band results to be ungapped,  
at least in the explored spectral range.
Our results show that only a two-band model can 
coherently describe both normal and superconducting state data.
\end{abstract}

\maketitle

\section {Introduction} 

Among the new iron based superconductors (IBS), the  FeSe system, 
in spite of its rather low 
superconducting transition temperature (T$_c$=8 K), 
is of interest because of its simple structure with stacking 
of FeSe$_4$ tetrahedra layers, without the intermediate charge reservoirs 
{\cite{ref1}, which are present between the FeAs$_4$ 
layers in iron pnictides. 
Furthemore, T$_c$ can increase up to 14 K by doping with 
tellurium {\cite{ref2}, and 
up to 37 K under pressure \cite{ref3}. 
Finally, of particular interest is the 
finding that T$_c$ can increase up to 65 K in a single 
layered FeSe film on SrTiO$_3$ \cite{ref4}.
As to the possible superconducting energy gaps in FeSe$_{1-x}$Te$_x$ 
systems, several experimental reports on their number, symmetry and 
value have not provided unambiguous results \cite{ref5,ref6,singh,konno}, 
presumably owing to 
differences in sample quality, disorder effects, and to the different 
sensitivity of the various experimental techniques to different gaps.
On the above basis, further studies on different
FeSe$_{1-x}$Te$_x$ systems are mandatory. 

In the present work we report an infrared spectroscopy study of 
a FeSe$_{0.5}$Te$_{0.5}$ film, aimed at the study of the low
energy electrodynamics of this chalcogenide system in both the normal
and superconducting states. Infrared spectroscopy is indeed a powerful 
technique to investigate both charge dynamics and band structure as it 
probes both free carriers and interband excitations, thus
providing a separation of the various contributions to the 
frequency dependent conductivity \cite{wooten,dressel}. 
In the superconducting state, in particular, mesurements performed in 
the THz region can provide direct information about formation
and presence of energy gaps \cite{tinkham,dressel}. 
In the case of pnictides, a large number of infrared/THz studies has
been reported; for example, in the case of the widely studied 
Co-doped BaFe$_2As_2$ system, 
more than 15 works have been published, as
reported in Ref.\cite{perucchi13}. 
On the contrary, in the case of FeSe$_{1-x}$Te$_x$ systems, the number 
of infrared studies is until now rather limited
(see the recent papers \cite{chinphys,pimenov} and references therein), 
owing presumably to the lack of high quality 
samples suitable for infrared/THz experiments. 
Therefore, a general agreement on the model conductivity and on 
number, symmetry and value of the possible superconducting gaps has not been 
achieved.

\section{Experimental}

The film under investigation is  a FeSe$_{0.5}$Te$_{0.5}$ 
film 200 nm thick 
deposited on a single crystal LaAlO$_3$ (001) substrate
by an ultra high vacuum pulsed laser deposition (PLD).
More details on sample preparation and 
morphology can be found in Ref.\cite{genova1,genova2,genova3}.
Note that the T$_c$ of these samples (13.7 K in the 
present case) can be as high as 20 K, but strongly depends on 
film thickness, because of strain effects, and on the growth 
conditions, in particular on the growth rate \cite{genova2}.
The 20 K residual resistivity $\rho_0$ of the film investigated 
is close to 1 m$\Omega$ cm, as shown in the resistivity curve 
$\rho$(T) reported in inset (a) of Fig.\ref{fig1}.
It is worth to recall that Te doping in FeSe samples can give origin 
to significant nonstoichiometry, disorder and clustering phenomena.  
In the case of films, these effects might lead to a higher 
defect density in the structure and thus to a lower DC 
conductivity compared to that of more ordered 
FeSe$_{1-x}$Te$_{x}$ single crystal and FeSe film samples.
The  presence of these defects is also confirmed by the superior 
pinning properties showed by PLD films in transport properties 
and by their direct observation by high resolution transmission 
electron microscopy \cite{genova4}.  

The absolute reflectance $R(\omega)$ has been measured 
with a Bruker 70v interferometer in the 80-8000 cm$^{-1}$ range
\footnote{Useful conversion factors are: 
1 eV=8066 cm$^{-1}$, 1 THz=33.4 cm$^{-1}$,
1 $\Omega^{-1}$cm$^{-1}$=4.78 cm$^{-1}$}
by using a gold mirror as reference and 
various beamsplitters, detectors and thermal sources. 
The 20 K $R(\omega)$ spectrum is reported up to 900 cm$^{-1}$ 
in Fig. \ref{fig1}. In inset (b), the reflectance is shown 
in the full range together with that of LaAlO$_3$ \cite{calvaniLAO}.
It is evident that the spectrum of the 
film+substrate system is dominated by the reflectivity of the 
substrate, as expected in the case of a low conductivity film 
deposited on a highly reflecting substrate. Measurements
performed on varying temperature from 300 to 20 K
did not show remarkable effects of the temperature on the measured 
spectrum. 

\begin{figure}
\includegraphics[width=8.8 cm]{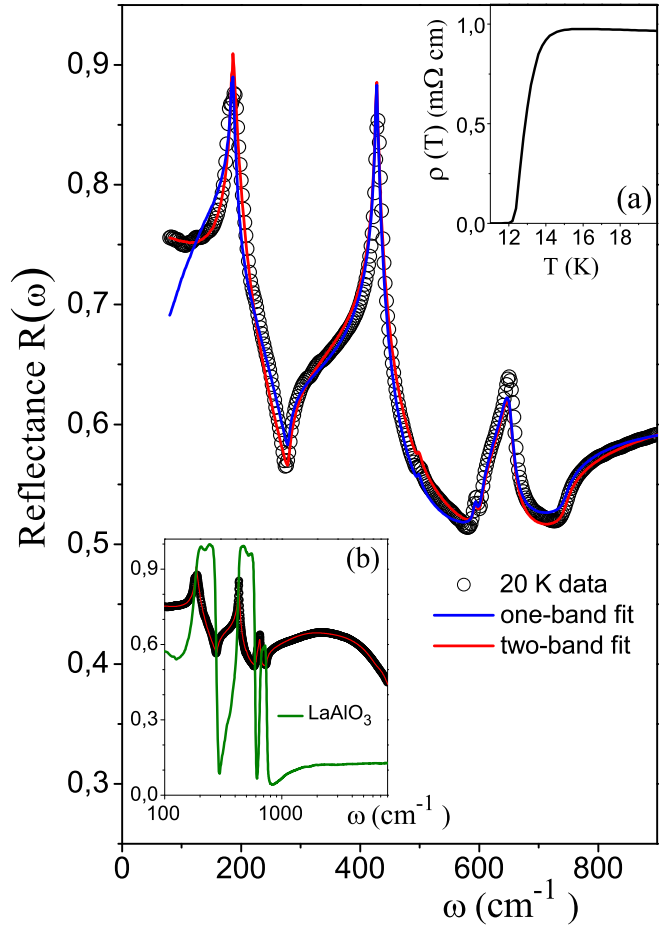}
\caption{Color online. 
Absolute reflectance $R(\omega$) of the 
FeSe$_{0.5}$Te$_{0.5}$ film at 20 K. Best fit curves according 
to the one-band and two-band models (see text) are reported.
Inset (a): resistivity curve $\rho$(T). Inset (b): Full range $R(\omega$) 
and two-band fit in log scale, compared with that of a bare  
LaAlO$_3$ substrate \cite{calvaniLAO}. } 
\label{fig1}
\end{figure}

THz measurements in the superconducting state were made by using 
synchrotron radiation from the SISSI beamline \cite{lupi07}
of the Elettra synchrotron; the interferometer was equipped
with thick mylar beam splitters and a 1.6 K liquid helium bolometer.
We measured the reflectance ratios $R(T)/R_N$ 
(with $R_N$ being the 20 K normal state reflectance) by cycling 
the temperature up to 20 K and below T$_c$ down to 6 K, without 
moving the sample. 
This technique does indeed provide results which are intrinsically 
unaffected by possible misalignments between sample and reference, 
and is thus essential in detecting the often small effects due to 
the superconducting transition \cite{palmer68,perucv3si,perucchi13,xi,xi2}.

Repeated measurements of the reflectance $R(T)$ at the 
minimum temperature 6 K ($R_S$), just above T$_c$ at 15 K,
and in the normal state at 20 K ($R_N$), provided  
reflectance ratios in the range 12-160 cm$^{-1}$ 
($\simeq$0.35-4.85 THz), 
as reported in Fig. \ref{fig2}a. We verified that, 
on increasing temperature, the $R(T)/R_N$ ratio progressively 
flattens, as expected. We remark that the use 
of a high-flux synchrotron source and 
the large size of the film surface allow high accuracy data  
like those in Fig. \ref{fig2}, extending down to a very low 
frequency.   

\begin{figure}
\includegraphics[width=8.8 cm]{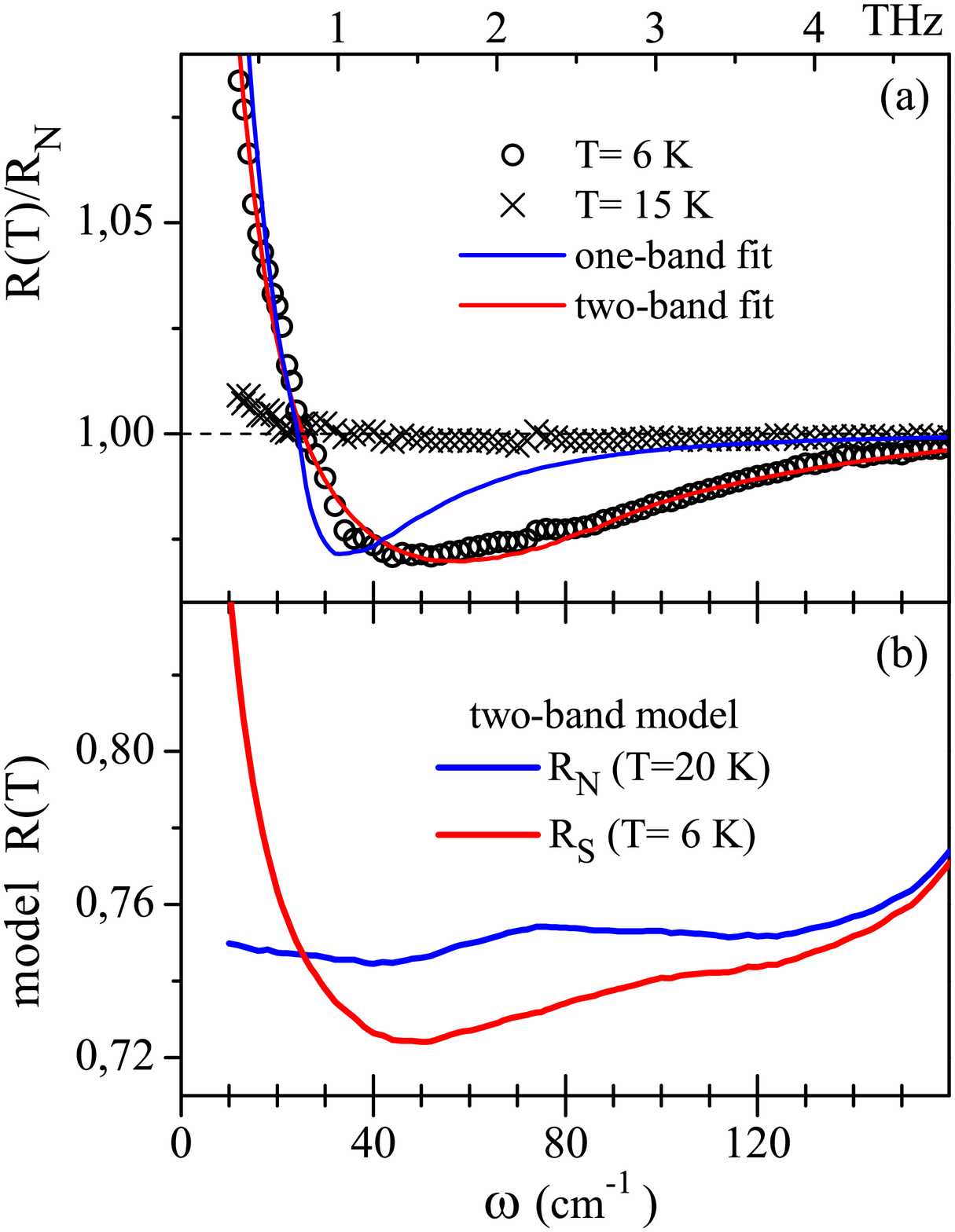}
\caption{Color online. (a) R(T)/R$_N$ reflectance ratio at 15 K and 
at 6 K (R$_S$/R$_N$). 
The best fit curves of R$_S$/R$_N$ by using one-band and
two-band models are shown (see text). 
(b) model R$_S$ and R$_N$ spectra given by two-band best fit results}  
\label{fig2}
\end{figure}

\section{Analysis and discussion}
The presence of several bands crossing the Fermi level is predicted in IBS.
Indeed, both the FeAs layer in pnictides and the FeSe layers 
in chalcogenides have holelike Fermi pocket at the zone center and 
electronlike Fermi pockets at the zone corners. 
Therefore, a minimal model can include one electron and one hole band,
which give origin to two different Drude contributions in the system 
conductivity. 
However, we want to note that, for related systems, the 
possibility of orbital differentiation has been recently reported as a 
possible origin of different Drude bands \cite{zhe,yi}.

As to the superconductin state, in a two-band scenario, it has been shown that 
both superconducting gap functions have $s$-wave symmetry, but with a possible 
sign change between the two ($s_{\pm}$ symmetry state)
 \cite{schachinger,hirschfeld}.
Experimentally, as reviewed in a recent work \cite{konno}, two gaps 
were observed  in a number of FeSe$_{1-x}$Te$_{x}$ samples
by using different techniques. For 0.5$<$x$<0.6$, 
a small gap was quoted at an energy as low as 0.51 meV \cite{bendele}, a 
large gap at an energy as high as 5.08 meV \cite{homes}. 

As to optical investigations, a two-band model was often adopted to describe 
the optical properties of different IBS systems (see for example
Refs. \cite{kim10,wu10,vanheu10} in the case of Co-doped BaFe$_2$As$_2$). 
To analyze the normal state spectrum we thus described 
the complex dielectric function 
$\tilde {\epsilon}(\omega)=\epsilon_1(\omega)+ i \epsilon_2(\omega)$
by employing the standard Drude-Lorentz model \cite{wooten,dressel}:
\begin{equation}
\tilde{\epsilon}=\epsilon_1(\omega) + i\epsilon_2(\omega) =  \epsilon_\infty-
\frac{\Omega_A^2}{\omega^2+i\omega\Gamma_A} -
\frac{\Omega_B^2}{\omega^2+i\omega\Gamma_B}+
\sum_j \frac{S_j^2}{{\omega_j}^2-\omega^2-i\omega\gamma_j} 
\label{lorentz}
\end{equation} 
where $\epsilon_\infty$ is the real part of the dielectric function at high 
frequency. $\Omega_A$ ($\Omega_B$) and $\Gamma_A$ ($\Gamma_B$) are
plasma frequency and scattering rate for the A (B) Drude contribution; 
$\omega_j$, $\gamma_j$ and $S_j$ are position,
width and strength of the $j-th$ Lorentz term accounting for a possible 
interband transition.
Note that the two Drude components are simply added according to 
a parallel conductivity model \cite{ortolani08}.   
We briefly recall \cite{wooten,dressel} that 
$\tilde {\epsilon}(\omega)$ directly provides the complex 
refractive index $\tilde{n}(\omega)=n(\omega)+ik(\omega)$,
which determines the optical response, and the complex conductivity     
$\tilde{\sigma}(\omega)= \sigma_1(\omega)+ i \sigma_2(\omega)$=
$i\omega$[$\tilde {\epsilon}(\omega)-\epsilon_\infty]/4\pi$, which is 
usually introduced in discussing the low energy electrodynamics
of the system. Here it is also worth to notice that, for a given Drude 
component, the d.c. conductivity $\sigma_0$=$\Omega^2$/$\Gamma$ is given by the 
zero-frequency value of the optical conductivity $\sigma_1(\omega)$, 
and that $\Omega^2$ is proportional to the integral of $\sigma_1(\omega)$,
$i.e.$ to its spectral weight. 

The model spectrum we employed to fit the experimental data is 
evaluated by using an ideal film+substrate system in which,
besides the finite thickness of the film, 
possible reflections from the substrate via Fresnel 
equations \cite{noiMedia,dressel} are taken into account. 
Detailed data on the LaAlO$_3$ complex refractive 
index \cite{calvaniLAO, noiTlao} are used.
The model spectrum thus only depends on the parameters 
in Eq. (1). We find that both models in which one Drude or two Drude 
terms are included (one-band and two-band model, respectively) 
can be used to describe the 20 K $R(\omega)$ spectrum, 
with similar accuracy.

In order to resolve such ambiguity, we turned to the study of the 
R$_S$/R$_N$ spectrum.
Here, we use  the procedures successfully employed in the analysis 
of the THz response of different superconductors such as 
MgB$_2$ \cite{ortolani08}, V$_3$Si \cite{perucv3si}, 
and Co-doped Ba122 films \cite{perucchi10,perucchi13}. 
In this procedure, each Drude term 
can be substituted by the Zimmermann term \cite{zimmermann91}, 
thus introducing two new parameters (superconducting gaps 
$\Delta_A$ and $\Delta_B$) in the fitting procedure. 
Note that the Zimmermann 
model describes the electrodynamics of a superconductor with
arbitrary gap $\Delta$ and relaxation rate $\Gamma$, thus generalizing 
the standard BCS Mattis-Bardeen model \cite{tinkham,dressel}.

Model spectra simultaneously fitted to both the R$_N$ and 
the R$_S$/R$_N$ spectra clearly show that good fits of both 
R$_N$ and R$_S$/R$_N$ in the THz region can be obtained only in the case 
of the two-band model}, as shown in Fig.\ref{fig1} and Fig.\ref{fig2}a.
In Fig.\ref{fig2}b we report the two-band R$_S$ and R$_N$ model 
spectra given by best fit results, to highlight the effect of the 
superconducting transition on the model reflectance spectrum.  
We remark that in the employed procedure, 
the simultaneous fitting of R$_N$ 
and R$_S$/R$_N$ imposes strong constrains to the parameter values. 
On the contrary, fitting procedures taking into account R$_N$ and 
R$_S$/R$_N$ separately, do not provide an univocal determination 
of the relevant parameters.

Our main findings are thus the following. In the first place, 
only a two-band model conductivity can describe all the experimental 
data, with one Drude component (A) taking up most 
of the spectral weight, being much broader than the other (B). 
It is worth to recall that a similar model conductivity 
has been adopted to describe the optical properties of different 
Co-doped BaFe$_2$As$_2$ systems \cite{lucarelli,nakajima,perucchi13}.
Secondly, a superconducting gap clearly opens up in the 
narrow (B) Drude band 
($\Delta_B$=37$\pm$3 cm$^{-1}$ $\simeq$4.6$\pm$0.4 meV), 
while the no gap signature in the broad (A) Drude band is detectable
within the experimental accuracy in the explored spectral range.
We remark that the presence of a small low energy gap, like that revealed 
at 0.51 meV$\simeq$4 cm$^{-1}$ from $\mu$SR \cite{bendele}, cannot be ruled 
out on the basis of present data extending down to 12 cm$^{-1}$.
The optical conductivity $\sigma_1(\omega)$ 
of the two bands in the normal state ($\sigma_{1NA}$ and 
$\sigma_{1NB}$) is plotted in Fig.\ref{fig3}a. 
Also the $\sigma_1(\omega)$ of the B-band 
in the superconducting state ($\sigma_{1SB}$) is plotted,
in order to point out the effect of the optical gap at 2$\Delta_B$.

It is worth reminding here that the reflectance measurements 
reported by Homes {\it et al.} \cite{homes} on a FeSe$_{0.45}$Te$_{0.55}$ 
single crystal in a wide frequency range (2 meV-3.5 eV) and  
the transmission experiments in the sub-terahertz region 
(5-35 cm$^{-1}$) reported by Pimenov {\it et al.} \cite{pimenov} 
on a FeSe$_{0.5}$Te$_{0.5}$ film, were well accounted for by 
using a one-band model conductivity, at least well above T$_c$. 
On the contrary, in both papers a two-band, 
two-gap model was adopted in the superconducting state.
We also remind that in the case of the reflectance measurements 
made for T$>$T$_c$ on a FeSe$_{0.3}$Te$_{0.7}$ \cite{mirri} single crystal, 
two Drude components were employed to describe obtained results, 
at least well above T$_c$.   
In this non well defined scenario, we remark that our analysis is 
the first being based on a coherent modelling of results obtained 
in both normal and superconducting state.

As to the normal state parameters, we obtain $\Omega_A$=7050 cm$^{-1}$ 
and $\Gamma_A$=1050 cm$^{-1}$, and thus 
$\sigma_{0A}$=790 $\Omega^{-1}$cm$^{-1}$ for the broad A-band. 
For the narrow B-band, $\Omega_B$=1600 cm$^{-1}$,  
$\Gamma_B$=200 cm$^{-1}$, $\sigma_{0B}$=220 $\Omega^{-1}$cm$^{-1}$.
Consequently, the total $d.c.$ conductivity 
$\sigma_{0tot}$=$\sigma_{0A}$+$\sigma_{0B}$ is close to
1000 $\Omega^{-1}$cm$^{-1}$, corresponding to a normal state 
resistivity  $\rho_0$=1/$\sigma_{0tot}$ close to 1 m$\Omega$ cm, in 
good agreement with the measured residual resistivity value 
(see inset (a) of Fig.\ref{fig1}). 

\begin{figure}
\includegraphics[width=8.8 cm]{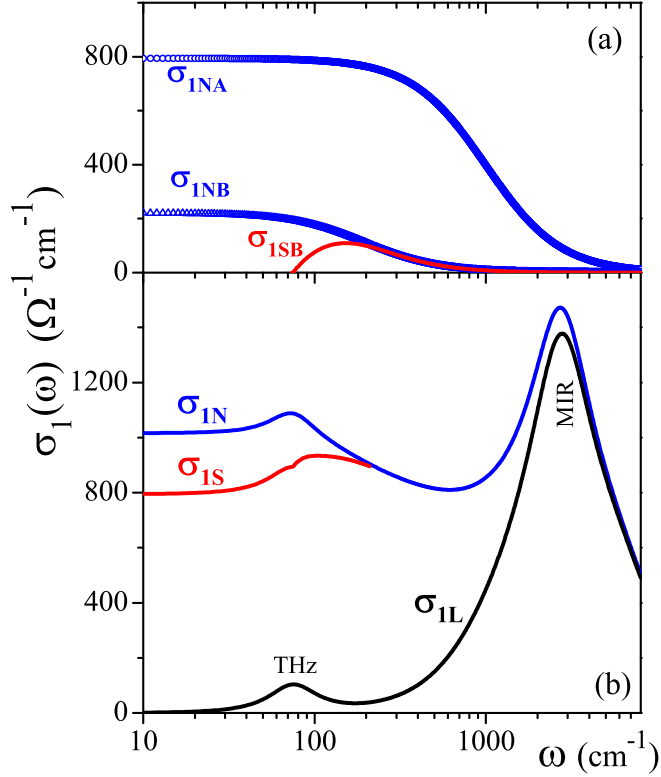}
\caption{Color online. 
(a) Optical conductivity $\sigma_1(\omega)$ of the A- and B-Drude 
bands in the normal state ($\sigma_{1NA}$ and $\sigma_{1NB}$). 
In the superconducting state, the effect of the gap opening in 
the B-band is shown in $\sigma_{1SB}$.
(b) Total $\sigma_1(\omega)$ in the normal ($\sigma_{1N}$) and
superconducting ($\sigma_{1S}$) state. In the contribution 
$\sigma_{1L}$ given by lorentzian terms, which is shown separately, 
the THz and MIR components are pointed out. 
}
\label{fig3}
\end{figure}

The total $\sigma_1(\omega)$ in the normal ($\sigma_{1N}$) and 
superconducting ($\sigma_{1S}$) states are shown in Fig.\ref{fig3}b. 
The contribution $\sigma_{1L}$
of the lorentzian terms which were included in the fitting 
procedure shows an intense mid-infrared component
(MIR in Fig.\ref{fig3}b) at about 
2500 cm$^{-1}$, which may be interpreted in terms of intersite 
electronic transitions between Fe atoms \cite{mirri}. 
The component located in the THz region at about 75 cm$^{-1}$
(THz in Fig.\ref{fig3}b) is analoguos to the one observed in Co-doped 
BaFe$_2$As$_2$ systems \cite{vanheu10,lobo,perucchi13}), 
which was interpreted as an intrinsic excitation \cite{benfatto} 
due to the onset of low-energy interband transitions.

The validity of the two-band model employed in our analysis is 
supported by a comparison of our normal state results 
with those obtained by Yuan et al. \cite{yuan} from the 8 K 
reflectance spectrum of a FeSe epitaxial film with T$_c$=7 K.
In the FeSe film, a broad Drude band (1) takes up most of the 
spectral weight, with respect to a second narrower, term (2) 
($\Omega_1$=10600 cm$^{-1}$, $\Omega_2$=2700 cm$^{-1}$). 
Therefore, the overall plasma frequency 
$\Omega_{tot}=\sqrt{\Omega_1^2+\Omega_2^2}$ 
results to be close to 
11000 cm$^{-1}$ \cite{yuan}, while in the present case 
$\Omega_{tot}\simeq$7250 cm$^{-1}$.
The discrepancy in the plasma frequency values can be explained 
by the better conductivity of the FeSe film of Ref. \cite{yuan} 
with respect to our FeSe$_{0.5}$Te$_{0.5}$ film, as noted in Sect.2. 
On the other hand, the ratios  $\Omega_B^2$/$\Omega_{tot}^2$, providing 
the spectral weight of the narrow Drude component with respect to the 
total, are in excellent agreement: 0.06 for the FeSe film from 
Ref.\cite{yuan}, and 0.05 for our FeSe$_{0.5}$Te$_{0.5}$ film. 
This result corroborates the validity of the two-band model 
conductivity approach.

As to the value of the superconducting gaps, our $\Delta_B$ estimate 
(37$\pm$3 cm$^{-1}$) is in very good agreement with that obtained in a 
strained FeSe$_{0.5}$Te$_{0.5}$ film 
($\Delta$=33$\pm$3 cm$^{-1}$) \cite{bonavol} by using a femtosecond 
spectroscopy technique of the quasiparticles relaxation times.
Note that this film and the one we investigated were 
prepared under the same conditions, in the same laboratory. 
As to the results obtained on different samples, 
Homes {\it et al.} \cite{homes} 
find that well below T$_c$ the optical conductivity 
of a FeSe$_{0.45}$Te$_{0.55}$ single crystal is well modeled in
the framework of a two-band, two-gap model with 
$\Delta_1$= 20 cm$^{-1}$ and $\Delta_2$= 41 cm$^{-1}$.
This result suggests either an $s_{\pm}$ or a nodeless extended 
$s$-wave gap functions. 
Also the conductivity of the FeSe$_{0.5}$Te$_{0.5}$ film investigated by
Pimenov {\it et al.} \cite{pimenov} in the sub-terahertz region 
was discussed below T$_c$ within the framework of a two-gap model.
In particular, the observed conductivity, besides showing weak features 
consistent with a low energy gap close to 12 cm$^{-1}$, 
was not completely suppressed below the corresponding optical gap, 
indicating the presence of a strongly anisotropic gap 
or nodes in the gap. Our model conductivity, in which the A-band 
is ungapped, at least in the explored frequency range,
might also well be consistent with this scenario.

\section{Conclusions}

We have reported an infrared spectroscopy study  of a 
FeSe$_{0.5}$Te$_{0.5}$ film with T$_c$=13.7 K, aimed at the study 
of the low energy electrodynamics of this chalcogenide system in both 
the normal and superconducting states. 
We analyze the 20 K normal state absolute reflectance R$_N$
measured over the broad infrared range 80-8000 cm$^{-1}$,
and the reflectance ratio R$_S$/R$_N$, 
R$_S$ being the superconducting state reflectance, 
measured at 6 K in the terahertz range down to 12 cm$^{-1}$. 
We find that only a procedure in which both spectra are simultaneously 
fitted poses strong constrains to the model parameters, and can thus 
provide an univocal determination of the model conductivity. 
Similar to previous results obtained for pnictide systems, 
a two-band Drude model is needed, where  one broad component provides 
the majority of the low-energy spectral weight.
In the superconducting state, we find that a gap opens up in the narrow 
Drude B-band 
($\Delta_B$=37$\pm$3 cm$^{-1}\simeq$4.6$\pm$0.4 meV), 
while the broad Drude A-band results to be ungapped,  
in the sense that no signature of a $\Delta_A$ gap is evident in the 
explored spectral range.

On the basis of a comparison with literature data, showing that there 
is still no general agreement between different reports, 
we believe that it is not possible to draw a definite conclusion on 
the presence of residual conductivity at finite frequencies in 
the superconducting state, as well as on number and symmetry of 
possible superconducting gaps. Further investigations on high quality 
Te-doped FeSe systems, possibly combining different experimental 
techniques, are thus mandatory. 
In particular, since inhomogeneities have been observed in these 
systems on a nanometer length scale \cite{bjoseph,genova4,singh}, 
it will be important to investigate their possible effect on the optical
response of a macroscopic sample portion, like that probed by an 
infrared beam.

\section*{Acknowledgments}
The work in Trieste was partially supported by Italian Ministry
of Research (MIUR) program FIRB Futuro in Ricerca
grant no. RBFR10PSK4. Work in Rome and Genova
was partially supported by the MIUR PRIN2012 Project
No.2012X3YFZ2.

\end{document}